\documentclass[10pt, numbers]{sigplanconf}
\pdfoutput=1
\usepackage{etex}

\usepackage[2ndsubmission]{oopsla2016}

\usepackage{etoolbox}
\makeatletter
\patchcmd{\maketitle}{\@copyrightspace}{}{}{}
\makeatother

\usepackage[utf8]{inputenc}

\usepackage{balance}
\usepackage{mathptmx}
\usepackage[english]{babel}
\usepackage{multirow}
\usepackage{rotating}
\usepackage{color}
\usepackage{xspace}
\usepackage{tabularx}
\usepackage{balance}
\usepackage{caption}
\usepackage{framed}
\usepackage{mdframed}
\usepackage{latexsym}
\usepackage{amssymb}
\usepackage{amsmath}
\usepackage{subcaption} 
\usepackage[inline]{enumitem}
\usepackage{xcolor}
\usepackage{algorithm}
\usepackage{algorithmic}

\usepackage{listings} 
\usepackage{textcomp}
 \usepackage[colorlinks,linkcolor=blue,citecolor=blue,urlcolor=blue]{hyperref}

\definecolor{primary}{RGB}{117,112,179}
\definecolor{secondary}{RGB}{27,158,119}
\definecolor{tertiary}{RGB}{217,95,2}

\definecolor{orange}{RGB}{255,127,0}
\definecolor{grey}{RGB}{135,135,135}

\AtBeginDocument{

}

\usepackage[]{pdfcomment}
\newcommand{\TODO}[1]{\textcolor{red}{#1}}\newcommand\todo\TODO

\newcommand{\question}[2]{{\bf RQ#1. #2}}
\newcommand{\answer}[2]{\vspace{.0cm}\noindent\fbox{\parbox{0.45\textwidth}{\textbf{Answer to RQ#1}. #2}}}

\newcommand{\tabincell}[2]{\begin{tabular}{@{}#1@{}}#2\end{tabular}}

\newcommand\banditrepair{BanditRepair\xspace}
\newcommand\exploitationCoefficient{$\zeta$\xspace}

\newcommand\nbBugs{16\xspace}
\newcommand\nbLaps{200\xspace}

\newcommand\stratReplaceVar{S1\xspace}

\newcommand\stratReplaceNew{S2\xspace}

\newcommand\stratSkipLine{S3\xspace}
\newcommand\stratReturnNull{S4a\xspace}
\newcommand\stratReturnVar{S4c\xspace}
\newcommand\stratReturnNew{S4b\xspace}

\newcommand\nbValidRuntimePatches{8460\xspace}

\newcommand\thetitle{BanditRepair: Speculative Exploration of Runtime Patches}

\copyrightdata{978-1-nnnn-nnnn-n/yy/mm} 
\publicationrights{author-pays}

\begin{document}
\title{\thetitle}

\authorinfo{Thomas Durieux}
           {University of Lille \& INRIA, France}
           {thomas.durieux@inria.fr}
\authorinfo{Youssef Hamadi}
           {Ecole Polytechnique, LIX, France} 
           {youssefh@lix.polytechnique.fr}
\authorinfo{Martin Monperrus}
           {University of Lille \& INRIA, France}
           {martin.monperrus@univ-lille1.fr}

\maketitle

\begin{abstract}

We propose, \banditrepair, a system that systematically explores and assesses a set of possible runtime patches.
The system is grounded on so-called ``bandit algorithms'', that are online machine learning algorithms, designed for  constantly balancing exploitation and exploration.
\banditrepair's runtime patches are based on modifying the execution state for repairing null dereferences.
\banditrepair constantly trades the ratio of automatically handled failures for searching for new runtime patches and vice versa.
We evaluate the system with \nbBugs null dereference field bugs, where \banditrepair identifies a total of \nbValidRuntimePatches different runtime patches, which are composed of 1 up to 8 decisions (execution modifications) taken in a row. 
We are the first  to finely characterize the search space and the outcomes of runtime repair based on execution modification. 
\end{abstract}

\section{Introduction}
Field failures happen in production for any software system of sufficient complexity. For instance, it is common to observe error pages on the Internet while ordering a laser pointer, registering to a conference, or installing a new blogging platform. Many of them have an economic cost and the most dramatic software failures lead to loss of lives.
To overcome software failures at runtime, runtime repair techniques modify the execution so that failures become less critical: instead of crashing the whole system, only the current task fails and the system remains available \cite{rinard2004enhancing,dobolyi2008changing,long2014automatic}. The literature refers to those modifications as ``runtime patches'' \cite{novark2007exterminator,Berger2012,zhang2014appsealer}. For instance, with failure oblivious computing \cite{rinard2004enhancing}, a runtime patch consists of skipping erroneous writes out of an array's bounds. In probabilistic memory safety \cite{berger2006diehard}, the runtime patches are controlled blank padding added around allocated memory. In the latter case, the runtime patch is a preventive measure and the execution is equivalent with or without the runtime modification. However, in the failure-oblivious case, the runtime patch has modified the system state or execution flow in an irreversible way. 

In this paper, we consider the case where multiple runtime patches exist to repair the same failure, which is a scenario that has been very little studied \cite{kling2012bolt,long2014automatic}. For instance, for repairing a null dereference, one can skip the execution of the statement or craft an arbitrary value before dereferencing \cite{long2014automatic}. 
We propose, \banditrepair, a system that systematically explores and assesses a set of possible runtime patches. The system is grounded on so-called ``bandit algorithms'' \cite{white2012bandit}, that are machine learning algorithms developed for A/B testing, which is the art of identifying the best strategies in production with in-the-field controlled experiments.

\subsection{Bandit Algorithm for Repair}
A bandit algorithm has two goals: 1) to systematically explore the search space of alternatives (e.g. all possible orderings of products) and 2) to
maximize the sum of rewards earned by successively trying alternatives (e.g. selling as many products as possible). 
In a runtime repair context, we instantiate the ``bandit view'' as follows: \banditrepair 1) systematically explores the search space of runtime patches and 2) increases the number of failures handled by the application of a runtime patch, where handled refers to corralling as follows.
A runtime patch is considered to handle\footnote{or ``repair'' or ``recover'', all being considered exchangeable in this paper} a failure, if a failure is replaced by the absence of exceptions within the scope of a given task, such as web request; if domain-specific post-recovery assertions exist, they are considered as well.

The key of bandit algorithms is to constantly balance exploitation (e.g. choosing the best product ordering seen so far) and exploration (e.g. finding an even better product ordering).
\banditrepair is configured by an exploitation coefficient \exploitationCoefficient ($\in ]0,1]$), that constantly steers the trade-off between the ratio of handled failures and the search for new runtime patches.
In \banditrepair, the exploration of runtime patches is speculative in the sense that one never knows in advance whether a state or flow modification is successful to handle a failure.

We implement \banditrepair for Java, in a version that is dedicated to null dereferences. To handle null dereferences, \banditrepair takes a decision in a pool of possible ones related to null pointer exceptions (creating objects, reusing objects, skipping execution). 
Sometimes, a failure requires multiple decisions to be taken in a row, which results in what we call a decision sequence.
A runtime patch is a decision sequence that handles a failure.
The simplest runtime patches are unary: they are composed of a single dereference repair decision, such as skipping the statement or returning from the method. An example of more complex runtime patch in \banditrepair is for instance: at line 24, a new object is crafted to overcome the null dereference, later on in the execution, at line 42, a statement is skipped to handle a second, subsequent null pointer exception. In such compoiste runtime patches, only one decision in isolation may not be enough to overcome the failure, only the sequence is a solution. More generally, a runtime patch contains decisions taken according to a ``runtime patch model''. To this extent, \banditrepair  is realized with a runtime patch model for null pointer exceptions.

\subsection{Pareto Front of Runtime Repair}

We evaluate \banditrepair on \nbBugs field failures reported for Java software. We run those field failures in a virtual endless ``while(true)'' loop that simulates the same failure happening again and again for different users, for different requests. By doing this, we can systematically study the search space of runtime repair, in terms of how many runtime patches exists, and how the exploration of alternative runtime patches can be balanced with exploitation of already known ones.

For instance, let us consider a field failure of Java open source package Apache Commons Collections, reported as issue \#360, which is about a null pointer exception.
We simulate 200 field failures by reproducing the failure 200 times in a row.
If one configures \banditrepair's exploitation coefficient to explore more than exploit, \banditrepair tries 38 decision sequences and finds that 15 of them overcome the null dereference.
On the other hand, by configuring \banditrepair to exploit known valid runtime patches more, \banditrepair tries 27 decision sequences and finds that 13 of them are valid solutions.
In the former cases, 62/200 failures are handled by 15 different runtime patches, in the latter case 154/200 failures are handled by 13 runtime patches. 
That is, we can construct the Pareto front of runtime repair, along two axes that are the number of different runtime patches identified (exploration) and the proportion of handled failures (exploitation). 

In our experiment, by summing over the \nbBugs considered null dereference field bugs, \banditrepair identifies a total of \nbValidRuntimePatches different runtime patches, which are composed of 1 up to 8 decisions (execution modifications) taken in a row.

\subsection{Reasons for Success}
\banditrepair  is capable of constructing the Pareto front of runtime repair for the following reasons.
First, it introduces a key concept for reasoning on modified execution states or flows, the one of ``execution laps''. 
An execution laps is a time-bounded logical unit of computation such as a web request or a command-line execution of a program. 
The execution laps is essential because:
1) it delimits the start and the end of a runtime patch: a runtime patch is composed of all execution modification decisions that happen within a laps. 
2) It comes with a predicate of the end of the laps that assesses the viability of the runtime patch. For instance, a web request may end with a HTTP success (200) or an internal server error (500), and a predicate may be "return\_code == 200". We call such predicate a ``laps oracle''.
By structuring speculative execution with those novel and original concepts, \banditrepair is able to systematically reason on the search space of runtime patches. 

Second, the fact that multiple runtime patches exist for the same failure is related to the deep nature of computation.
In a single program execution, there are parts of the computation that are optional with respect to the task at hand. If a failure happens in this optional part, it impedes the whole task \cite{ammann1988data,rinard2004enhancing}. If given enough time to explore the search space, \banditrepair automatically finds the execution shortcuts that exist to skip a failure.
The other fundamental characteristics of software exploited by \banditrepair is that there exists multiple execution paths to achieve the same computational effect. When those multiple paths are identified and a failure happens, one alternative path may succeed.   
\banditrepair automatically builds a portfolio of alternative execution paths in a systematic manner.

\subsection{Contributions}

To sum up, the contributions of this paper are:
\begin{itemize}
\item A runtime repair algorithm, \banditrepair, that balances exploitation and exploration of runtime patches. The algorithm is implemented in Java, for repairing null pointer dereferences, and is made publicly-available for sake of open science.
\item The characterization and systematic empirical study of the repair search space for null dereferences with respect to:
\emph{(Size)} how many different repair decisions can be tried to handle a given failure?;
\emph{(Fertility)} how many valid sequences of repair decisions exist in the search space?;
\emph{(Disparity)} are all repair decisions equal?;
\emph{(Trade-off)} what is the impact on the exploitation/exploration balance on search space exploration.
\item An evaluation over \nbBugs null dereference failres, reported in the field on a public issue tracker, on largely used Java libraries. By simulating the perpetual occurrences of those field failures, the evaluation identifies \nbValidRuntimePatches valid runtime patches, in an experiment that represents more than 10 days of computation in a distributed grid.
\end{itemize}

The remainder of this paper is organized as follows.
\autoref{sec:concepts}  presents our approach for repairing null pointer exceptions at runtime.
\autoref{sec:contribution} details our \banditrepair algorithm.
\autoref{sec:evaluation} details the evaluation on \nbBugs field null dereference.
\autoref{sec:rw} presents the related works and Section \ref{sec:conclusion} concludes.

\section{Problem Statement}
\label{sec:concepts}

\subsection{Motivating Example}
Let us consider the example of \autoref{fig:npe-example}. 
It is an excerpt of server code that retrieves the last connection date of a user and prints it to an HTML page.
Method getLastConnectionDate first gets the user session, then pulls the last connection date from the session object.
This snippet can trigger two failures that can crash the request:
1) if the session does not exist and getUserSession returns null, then there is a null pointer exception at line 3 (NPE1)
2) for the first connection, getLastConnection returns null, and another null pointer exception can be thrown at line 6 (NPE2).
Now let us consider a runtime repair system such as \cite{DBLP:conf/pldi/LongSR14}. It would insert hooks in code such that, instead of a null dereference, a viable object is crafted upon failure.
In \autoref{fig:npe-example}, to overcome NPE1 at line 3, such a system could modify the execution state and flow in three ways:
1) it creates a new session object on the fly
2) it returns an arbitrary Date object such as the current date
3) it returns null.
As the example suggests, \emph{there are multiple possible state modifications for the same failure}.
However, not all such modifications are equivalent. For instance, if modification \#3 is applied, it triggers another failure NPE2, whereas solutions \#1 and \#2 do not further break the system state. This indicates that  \emph{not all state modifications are equivalent, some are best than others}.
In this paper, we devise a system that speculatively explores execution modifications in order to identify the better ones.

\begin{lstlisting}[numbers=left, caption={Code Excerpt with Two Potential Null Dereference Failures},label=fig:npe-example]
Date getLastConnectionDate() {
    Session session = getUserSession();
    return session.getLastConnection(); // NPE1
}
...
HTML.write(getLastConnectionDate().toString()); // NPE2
\end{lstlisting}

\subsection{Problem statement}
\label{sec:problem}

In this paper, we consider the problem of production failures.
Since programs run in heterogeneous environments, responding to a large number of unpredictable events and inputs, 
errors happen, and manually written error handlers fail to handle them in many situations \cite{subkraut2006automatically,gunawi2008eio}.
Consequently, failures in production happen on a daily basis, and crash reporting systems routinely collect enormous amounts of failure information: for instance the publicly available crash reporting system of Mozilla collects more than 30,000 crash reports per day for its Firefox browser \cite{khomh2011entropy}. 
We note that the same failure often happens multiple times, again and again, for different users, on different servers, etc. 

One solution to this problem is to change the program state or flow such that the failure does not happen or is mitigated and the program is able to proceed with execution. 
This is known as runtime repair \cite{Lewis2010} and state repair \cite{Monperrus2015}.  One example of such a runtime repair approach is by Demsky and Rinard \cite{demsky2003automatic}, who have proposed automatic restoration of invariants for coping with certain errors that are specific to data-structure.
The ideal repair system would transform a failure into a correct result. However, this idealized vision is not realistic and in practice, the goal of runtime repair is to corral failure propagation, to replace crashed systems with continued execution, i.e. to increase overall availability.  

Preliminary work on runtime repair suggests that for a single crashing failure, there are several possible repair decisions to be made \cite{kling2012bolt,long2014automatic,cornu:hal-01062969}.
For instance, let us a consider a division by zero, there are several possible repair decisions that could be made: one can divide by 1, or the result of the division can be an arbitrary value (0, 1, MAX\_VALUE or Not-a-number among others).
In other words, upon failure, there are multiple and different alterations of program state that can be made to handle it. 
One alteration may not be enough to continue the execution after a failure, and after the first alteration of the program state,  another failure may happen, triggering another alteration and so on. This is known as ``cascaded errors'' \cite{long2014automatic}. The principled handling and systematic study of ``cascaded errors'' and the corresponding cascaded repair decisions is an open and unexplored research field.

\begin{mdframed}
Problem: we aim at devising an architecture and a system that handles failures, by exploring cascaded runtime repair decisions (called ``decision sequences'') in a principled and effective way.
\end{mdframed}

\subsection{Exploration of Runtime Repair Decisions}
\label{sec:exploration_runtime_decisions}

Our key insight is that we can exploit the fact that the same failure happens again and again to explore alternative repair decisions. For a given failure, our idea is to record the execution modification decision and its eventual effectiveness and then to steer the new ones according to the past decisions. 

We propose a conceptual framework for this, it seeks to balance the reuse of past runtime decisions (this is called exploitation) that have been shown successful and the exploration of new runtime decisions. The terminology exploitation/exploration as well as the core algorithm is inspired from bandit algorithms. 

Bandit algorithms are machine learning algorithms that aim at maximizing the profit of a player in front of multiple slot machines (aka one armed bandit machines).
To maximize its profit, the player aims at identifying the machine that has the largest probability of winning.
However, these probabilities are unknown, hence the player has to estimate them.
After an initial number of trials, he has an estimation of all winning probabilities, with one being higher than the others.
However, it may only be due to the variance of the estimation process.
Consequently, he has to balance playing the slot machine with highest probability (exploitation) and playing the other machines to gain knowledge (exploration).

We think that the two opposing yet complementary concepts of exploitation and exploration perfectly fit the problem of runtime repair. When a failure happens, from which one knows that a solution exists (based on previous executions), one can either reuse the existing knowledge (apply the same repair decision) or explore a new solution, which may prove to be better. 
As shown in \autoref{fig:npe-example}, a runtime patch that does not trigger new failures is better than one that creates invalid program states.
The trade-off between exploration and exploitation of runtime repair is the essence of the system we present in this paper, a system called \banditrepair. \banditrepair is inspired from the so-called epsilon-greedy bandit algorithm \cite{white2012bandit}.

\section{\banditrepair : runtime repair of failures}
\label{sec:contribution}

In this paper, a {failure} is defined as an unacceptable interruption of service of a program for a given input. For instance, when a web server does not succeed to serve a file, it is a failure. When a command line program crashes, it is a failure.
Failures are considered deterministic: for the same input and the same system state the failure always or never happens.
We propose \banditrepair, a conceptual framework and an algorithm for automatically handling failures at runtime.

\subsection{\banditrepair Inputs}
\label{sec:input}

\banditrepair requires five inputs:
1) a program in which runtime repair support will be injected;
2) a failure model expressing the failures targeted by the system;
3) a laps model defining the boundaries between which runtime repair will take place;
4) a laps oracle specifying the viability of the computation;
5) a runtime patch model listing the possible modifications on the program state or execution flow.

\subsubsection{Program}\label{sec:program}
The first input of \banditrepair is a program $P$.
\banditrepair uses meta-programming to inject in the program a set of monitoring and runtime intercession hooks.
The monitoring hooks include failure detection according to the failure model (see \autoref{sec:failure_model}) and runtime contract checking according to the laps oracle (see \autoref{sec:laps_oracle}).
An example of failure detection is the detection of null dereference by automatically adding null check (``if (x!=null) {...}'') before all field accesses and method calls.
The runtime intercession hooks enable \banditrepair to change the program execution if appropriate according to the runtime patch model (see \autoref{sec:runtime-patch-model}).

\subsubsection{Failure Model}
\label{sec:failure_model}
\banditrepair is parametrized by a failure model. A \textbf{failure model} is an abstraction to represent a family of production failures. An example of failure models is divide-by-zero.
In \banditrepair, failure models are intentional definitions, where the necessary and  sufficient property is a \textbf{failure predicate}, which returns true if and only if the failure is about to happen. 
For detecting divide-by-zero errors, one can check upfront all denominators of divisions against zero.  
As explained in \autoref{sec:implementation}, in this paper, we realize \banditrepair for null dereferences (null pointer exceptions in Java).

\subsubsection{Laps Model}
\label{sec:laps}

A \textbf{laps model} defines a quantization of execution time.
Program execution contain many natural quanta: for instance a method execution is an execution quantum, this a method laps, a request handling in a web server is a quantum (a request laps), the full execution of a command line program is a quantum (a command-line laps).
As we see in those examples, a laps model defines when laps start and end.
\banditrepair works with any laps model that defines a laps start and end. 
Laps models can be domain specific, for instance, in a scientific simulation software application, a good laps model is a simulation step.

\subsubsection{Laps Oracle}
\label{sec:laps_oracle}
A \textbf{laps oracle} is a predicate on the program state that is executed at the end of each laps.
The goal of a laps oracle is to validate or invalidate state modifications that have happened during the laps. For instance, in a web-server with a request laps model, a laps oracle can be whether the HTTP request return code is OK (``assert response\_code == 200'').
While this example predicate only refers to one variable, it can be arbitrarily complex and refer to many parts of the observable program state.
If one considers method laps, a classical design-by-contract post-condition is a laps oracle.

A laps oracle can serve two purposes. First, it can assess whether what happened in the laps has not failed, as in the examples we have given. This is an assertion on the past. Second, a laps oracle can assess whether the system state is viable for future actions and requests, this is then an assertion looking ahead, for asserting what Locasto et al. have called ``life after self-healing'' \cite{locasto2008life}. Whether it assesses success or viability, the result of the evaluation of the laps oracle always tells something about the success of state modifications that have happened in the laps. In our empirical experiment, the laps oracles assess the validity of the past computation.

\subsubsection{Runtime Patch Model}
\label{sec:runtime-patch-model}

When a failure of the failure model under consideration is about to happen, as detected by the failure predicate, the program or request (depending on the laps model) is about to crash. It would indeed have crashed without \banditrepair. However, with \banditrepair, when a failure is about to crash the program, \banditrepair replaces the crash by a modification of the program state or flow. The modifications are done according to a runtime patch model. 

\textbf{Definition.} A \textbf{runtime patch model} describes how program state modifications are performed upon failures. 
Once a failure predicate evaluates to true, the program hits a decision point.
A \textbf{decision point} is a point in a program where the program state may be modified.
At a decision point, a decision must be taken: we call a \textbf{decision} an alteration of the execution state or flow.
For instance, prematurely returning from the method is such a modification. 
When a decision is taken, one says that the decision point has been activated.
In essence, a decision is speculative: \banditrepair never knows in advance whether the decision is correct.
The only way to assess it is to proceed with execution and wait for the evaluation of the laps oracle.

Within a laps, several decision points may be activated, due to cascaded failures.
A \textbf{decision sequence} is composed of consecutive decisions (runtime repair actions), where there is one decision per failure of the failure cascade. The decision are instances of the runtime patch model, the failures of the failure model.
At the end of a laps, the laps oracle is evaluated. If it evaluates to true, it means that a valid decision sequence has been found. 
A \textbf{runtime patch} is a decision sequence that has been validated by the laps oracle. For instance, let us consider again the example of \autoref{fig:npe-example}, returning null is a failed decision sequence because the request crashes with HTTP 500, internal server error, due to NPE1. 
On the contrary, returning a fresh date object enables the request to succeed and the HTML to be generated, this is a valid unary decision sequence, i.e. a runtime patch.

Within a runtime patch, the first activated decision point has a special status. First, it is where the program was about to crash: consequently, we call the first activated decision point in a laps the \textbf{failure point}.
Second, it is from there that the program execution will speculatively explore new runtime states.
In this paper, we implement \banditrepair with the ``NpeFix'' runtime patch model \cite{cornu:hal-01251960}, described in \autoref{sec:implementation}.

\begin{algorithm}[t]
  \begin{algorithmic}[1]
    
    \REQUIRE{P: a program}
    \REQUIRE{F: a failure detector}
    \REQUIRE{L: a laps model}
    \REQUIRE{O: a laps oracle}
    \REQUIRE{R: runtime patch model}
    \REQUIRE{\exploitationCoefficient: the exploitation coefficient in $[0..1]$}
    \ENSURE{S: a set of portfolio of runtime patches for each failure point location}
    \WHILE{true}
        \STATE{L.start()}
        \WHILE{failure $f$ happens according to F}
            \IF{failure is not known}
                \STATE{apply a random decision \bf{Case 1}}
            \ELSE 
                \IF{rand $<$ \exploitationCoefficient}
                    \STATE{apply up-to-now best runtime patch from S \bf{Case 2a}}
                \ELSE
                    \STATE{d $\leftarrow$ select an unused decision according to R}
                    \STATE{apply d (change the state or flow) \bf{Case 2b}}
                \ENDIF
            \ENDIF
            \STATE{proceed with laps execution}
        \ENDWHILE
        \STATE{L.end()}
        \IF{laps oracle $O$ is success}
          \STATE{store runtime patch in S}
        \ENDIF
    \ENDWHILE
  \end{algorithmic}
  \caption{The main algorithm of \banditrepair}
  \label{algo:top-level}
\end{algorithm}

\subsection{\banditrepair Effects}
\label{sec:decision-taking}

The core of \banditrepair is an engine that selects one decision when a failure is detected, that is when decision point is hit.

When a failure, instance of the failure model under consideration, is detected at a decision point, \banditrepair has to take one decision in order to handle the failure (instead of letting the program crash).
This is done as follows.

\textbf{Case 1:} The decision point has never been activated, which means that the failure has never happened in this location of the program before. For instance, when a null dereference has never been seen up to now at line 3 of \autoref{fig:npe-example}.
In this case, \banditrepair randomly selects a decision in the set of alternative possible decisions. 

\textbf{Case 2:} The failure has already been seen before at this point in the program. For instance, in a server program, it means that another user has already encountered the same failure, by performing the same sequence of interactions with the program. When this happens, \banditrepair has to choose between exploitation and exploration as follows.

\textbf{Case 2a exploitation:} When the failure has already happened, it means that one or more decisions have already been taken during another laps (i.e. the program execution has already been altered in the past in another request, for another user). For each past decisions, the laps oracle has been evaluated, and \banditrepair has stored whether the past failure was considered handled according to the laps oracle. When \banditrepair chooses exploitation, it selects one decision sequence which has been the most successful over the past laps. We use the term ``exploitation'' to refer to the fact that the system exploits its knowledge, by maximizing the likelihood of handling the failure.

\textbf{Case 2b exploration:} When the failure has already happened at a given decision point, \banditrepair may choose to take a decision that was never taken before at this point. In this case, \banditrepair speculatively explores new runtime patches. 

To choose between exploitation (case 2a) and exploration (case 2b) \banditrepair draws a random variable from a uniform distribution.
If it lower than an \textbf{exploitation coefficient} \exploitationCoefficient (zeta), it selects exploitation, otherwise it selects exploration.
For instance, \banditrepair with \exploitationCoefficient = 0.2 prefers exploitation 20\% of the time, and exploration 80\% of the time.

For large exploitation coefficients \exploitationCoefficient, \banditrepair often reuses known runtime patches, for low coefficients, \banditrepair faster explores the space of possible runtime patches. This effect will be empirically studied in \autoref{sec:evaluation}.

In the limit case of \exploitationCoefficient = 1, \banditrepair only performs case 1 and case 2a, which means that as long as one runtime patch is found at a decision point, it is always applied over and over. We call this the \textbf{full exploitation policy}.

\subsection{\banditrepair Algorithm}
\label{sec:core-algorithm}

Algorithm \autoref{algo:top-level} presents \banditrepair. It takes as input a program, a failure detector, a laps model, a laps oracle and a runtime patch model, as explained in \autoref{sec:input}.
Then, for every laps, if a failure is detected, \banditrepair randomly selects between exploitation (line 8) and exploration (line 10). If the laps oracle validates the decision sequence, it becomes a runtime patch (according to our definition of \autoref{sec:runtime-patch-model}) and is stored as such.
Overtime, \banditrepair builds a set of runtime patches for each failure location, we call it \textbf{a portfolio of runtime patches}, a portfolio of decision sequences that have proven successful at least once.

All runtime patches in a portfolio share the common property that they pass the laps oracle. However, they differ from two perspectives.
First, they may involve different decision points, at different locations in the program under consideration.
Second, they may have different size, where the size of the runtime patch is the number of decisions taken.
A runtime patch can be considered better if it contains fewer decisions, because it is likely to change less the execution state, and hence to stay closer to the states created by the initial program and envisioned by the developer. In other words, the smaller runtime patch creates execution states that are less speculative than the ones created by a bigger runtime patch.
This will be explored in \autoref{sec:rq3}.

\begin{table}
\caption{\banditrepair' possible decisions upon null dereference failures.}
\label{tab:strategies}
\centering
\resizebox{0.40\textwidth}{!}{
\begin{tabularx}{\columnwidth}{|l|l|l|l|X|}
\hline
\multicolumn{3}{|c|}{Strategy} &  Id & Description \\ \hline
\multicolumn{2}{|c|}{\multirow{3}{*}{replacement}}
 & reuse               & \stratReplaceVar   & injection of an existing compatible object \\ \cline{3-5}
\multicolumn{2}{|c|}{} & {creation}         & \stratReplaceNew  & injection of a new object \\ \hline
\multirow{7}{*}{\rotatebox{90}{skipping}} 
 & \multicolumn{2}{c|}{line}                & \stratSkipLine    & skip statement \\ \cline{2-5}
 & \multirow{6}{*}{\rotatebox{90}{method}} 
                                 & noting   & \stratReturnNull  & return a null or void to caller \\ \cline{3-5}
 &                               & creation & \stratReturnNew   & return a new object to caller \\ \cline{3-5}
 &                               & reuse    & \stratReturnVar   & return an existing compatible object to caller \\ \cline{3-5}
\hline
\end{tabularx}
}
\end{table}

\subsection{Implementation}
\label{sec:implementation}

We implement \banditrepair for  null dereferences, aka null pointer exceptions (this is the failure model) with a runtime patch model dedicated to them, called NpeFix \cite{cornu:hal-01251960}.
In NpeFix, all object variable dereferences are decision points (field accesses, method calls on local variables, method parameters, implicit casts and fields).
A decision has to be taken when the variable is null, which means that decision points are activated if and only if a null is going to be dereferenced.
For each decision point, NpeFix defines 6 types of decision, grouped in two categories, shown in \autoref{tab:strategies}. 
The first category consists of replacing the null value by an alternative valid non-null object of a compatible type. This category is composed of two-sub-categories:
1) when a variable is null, one can reuse an object from another variable in the scope instead, these are reuse-based decisions (on top of \autoref{tab:strategies});
2) when a variable is null, one can also create a new object on the fly, these are creation-based decisions. 
Note that the number of possible decisions for reuse and creation based decisions is parametrized by the number of variables (resp. constructors) available, which means that for a single decision point, there are often dozens of different available decisions (and not only 6).

The second category is based on skipping the execution of the code affected by the null variable, one can either skip the line that uses the null variable, or skip the rest of the method. When skipping the rest of a method which returns a value, one can also either reuse an existing object or create one on the fly. 
To implement the NpeFix runtime patch model, we use source code transformation: we inject code at each decision point and the injected code is responsible to activate the decision point and actually performing the state or flow modification if necessary.
For the interested reader, this runtime patch model and its implementation are extensively described in \cite{cornu:hal-01251960}.

\section{Evaluation} \label{sec:evaluation}

We now present the evaluation of \banditrepair.  
During this evaluation we focus on the following research questions.

\newcommand\rqCore{\question{1}{[Size] Does the core assumption of \banditrepair hold? How large is the runtime repair space?}\xspace}
\rqCore
Bandit exploration of runtime patches only makes sense under the following conditions: 
1) the repair space at runtime contains different alternatives;
and 
2) 
not all alternatives are valid.
The answer to this research question will (in)validate the core assumption of \banditrepair.

\newcommand\rqProportionValid{\question{2}{[Fertility] What is the proportion of valid decision sequences?}\xspace}
\rqProportionValid
In the context of runtime repair, there may exists different runtime patches that are all valid, that all fix the runtime failure. 
The proportion of valid repair decision sequences represents the fertility of the search space.
When the goal is to find at least one valid decision sequence, it is much easier to do so if many points in the search space are valid.
On the contrary, if there is a single point in the search space, in the worst case, it requires visiting the complete search space before finding it.
The fertility of the search space is the opposite of what is called ``hardness'' or ``constrainedness'' in combinatorial optimization.

\newcommand\rqBetterDecision{\question{3}{[Disparity] To what extent does the search space contain composite runtime patches?}\xspace}
\rqBetterDecision
\banditrepair builds a portfolio of runtime patches, which are disparate in the sense that they can have a different size. In this paper, we consider that a smaller runtime patch, i.e. containing fewer decisions, is better than a bigger composite one, because it creates less exotic execution states (see \autoref{sec:core-algorithm}). 
We will observe in our dataset, whether there exists such composite runtime patches.

\newcommand\rqExplorationCoefficient{\question{4}{[Trade-off] What is the impact of the exploitation coefficient \exploitationCoefficient on repair?}\xspace}
\rqExplorationCoefficient
The essence of bandit algorithms is to alternate exploitation of valid decisions and exploration of alternative ones.
In the context of runtime repair, it means applying a runtime patch that has proven to be successful or searching for alternative runtime patches. 
We will explore  the impact of the exploitation coefficient \exploitationCoefficient on the time to find a first runtime patch and the overall proportion of avoided failures.

\subsection{Dataset}\label{sec:dataset}
In order to evaluate our runtime repair approach, we need real and reproducible production failures. 
Since we instantiate the bandit repair vision with null pointer exceptions, we collect null dereference failures.

To collect them, we look for null dereferences that are reported on a publicly-available forum (e.g. a bug tracker) and we assess that they are reproducible.
In particular, we focus on failures in the Apache foundation projects because these projects are frequently used and have very good practices for bug reporting and field failure reproduction.  
In Apache, one guideline is to encode reproduced field bugs as test case. Consequently, our dataset of field bugs is composed of test cases, written by the developers of each project under consideration, which reproduce field bugs.
In addition to the triple criteria of being field, reproducible and encoded as test cases, we aim at 1) having bugs in different projects and 2) having bugs in large enough software (where ``large" is defined as more than 10,000 lines of code).

As a result, the benchmark contains \nbBugs field bugs (1 from Collections, 3 from Lang, 7 from Math, 3 from PDFBox, 1 from Sling and 1 from Felix).
This dataset only contains real null dereference bugs and no artificial or toy bugs.
To give the reader a feeling of how hard it is to reproduce field bugs, we note that it took us appropriately 1 full month to build this dataset.
As comparison, \cite{rinard2004enhancing} considers 5 field failures.
For sake of future work and comparative evaluations on this topic, this dataset is made publicly available on GitHub.\footnote{\banditrepair dataset: \url{https://goo.gl/937Egi}} 

\begin{table}[t]
\caption{Dataset of \nbBugs bugs with null dereference in six Apache open-source projects.}
\label{tab:dataset}
\centering
\resizebox{0.45\textwidth}{!}{
\setlength\tabcolsep{0.7 ex}
\begin{tabular}{|l|l|r|r|}\hline
Bug ID          & SVN revision& LOC   & \tabincell{c}{\# method calls \\ before null}\\\hline
Collections-360 & 1076034     & 21650 &                         13 \\
Felix-4960      & 1691137     & 33057 &                          2 \\
Lang-304        & 489749      & 17277 &                          2 \\
Lang-587        & 907102      & 17317 &                         10 \\
Lang-703        & 1142381     & 19047 &                          9 \\
Math-1115       & 1590254     & 90782 &                        328 \\
Math-1117       & 1590251     & 90794 &                        342 \\
Math-290        & 807923      & 38265 &                         88 \\
Math-305        & 885027      & 38893 &                          8 \\
Math-369        & 940565      & 41082 &                          7 \\
Math-988A       & 1488866     & 82442 &                        136 \\
Math-988B       & 1488866     & 82443 &                        134 \\
PDFBox-2812     & 1681643     & 67294 &                         37 \\
PDFBox-2965     & 1701905     & 64375 &                         54 \\
PDFBox-2995     & 1705415     & 64821 &                         37 \\
Sling-4982      & 1700424     & 1182  &                          2 \\
\hline
\multicolumn{2}{|l|}{Total: \nbBugs applications from 6 projects} & 770721  & 1209\\
\hline
\end{tabular}
}
\end{table}

\autoref{tab:dataset} presents our benchmark of \nbBugs field bugs.
The first column contains the Apache bug id.
The second column contains the SVN revision of the global Apache SVN.
The third column contains the number of line of code.
The fourth column contains the total number of method call before the null pointer exception is trigger.
For example, issue Collections-360 fixed at revision 1076034 is within an application 21650 lines of code.
The number of calls before the dereference gives an insight on the complexity of the setup required to reproduce the field failure. As shown in \autoref{tab:dataset}, there are between 2 and 342 application methods (not counting JDK methods) called for the reproduced field failures under consideration, with an average of 75.56. This indicates that the failures in our benchmark are not simple tests with a trivial call to a method with null parameters.  

\subsection{Experimental Protocol}
\label{sec:protocol}

We perform two experiments. 
The first one is based on the exhaustive exploration of the search space of runtime patches, as defined by our runtime patch model for null pointer exceptions described in \autoref{sec:implementation}.
The second experiment trades-off exploration and exploitation of the search space.
Both are done on the benchmark of failures presented in \autoref{sec:dataset}.

\subsubsection{Exhaustive exploration}
\label{sec:exhaustive-exploration}

To exhaustively explore the search space of runtime patches for a given failure, we simply recursively explore all possible alternative decisions.
For the first decision taken at the failure point (the first decision point in a laps), we take all decisions one after the other. 
Then, for each new decision points activated by the first decision, we also explore all possible decisions. This is done recursively.
In other words, we build the complete decision tree of repair decisions for a given failure. 

The time required to perform such an experiment has a bottom bound of the size of the search space multiplied by the time for reproducing the failure.
The alternative computation that comes after the first repair decision at the failure point is added on top of this.
Overall our experiment takes more than 10 days.

\subsubsection{Bandit exploration}
\label{sec:bandit-exploration}

The study of exploration of runtime patches is done as follows.
\begin{enumerate*}
  \item We instrument each buggy program of our dataset with our repair framework.
  \item We execute each instrumented program with the test case that encodes the field bugs.
  \item We collect all decisions taken at runtime
  \item We execute the runtime assertions at the end of the test cases 
\end{enumerate*}

We run step \#3 and \#4 a large number of times, it simulates users that trigger a production failure again and again. Indeed, production bugs keep reappearing as long as they are not fixed. This is why crash reporting systems have large number of instances of the same crash \cite{khomh2011entropy}.
We trigger all failures exactly \nbLaps times.
For instance, we run the crashing test case of bug LANG-304 \nbLaps times, simulating that the crash happens on \nbLaps user machines spread over the world, and communicating one another or to a server about the crashes, in an application community style \cite{Locasto2006}. 
In the following, a sequence of \nbLaps runs is called a scenario (the scenario of having \nbLaps users triggering the same failure). 

In addition, \banditrepair is parametrized by an exploration/exploitation coefficient. We would like to understand the impact of this coefficient. 
Consequently, we apply the whole process (step \#1 to step \#4) for 11 different exploitation coefficient \exploitationCoefficient from $0$ to $1$ with a $0.1$ step.
In the following, we use the term ``run'' to refer to one failure execution (one test case), for one given exploitation coefficient.

Finally, recall that our algorithm contains a random component. Our implementation fully controls this randomness by using a parametrized seed of the random number generator. However, it may happen that the system works accidentally well for a given seed.  
To mitigate this risk, for each exploitation coefficient, we repeat the process with 31 different random seeds.

In total, we execute $\nbBugs$ bugs $\times$ $\nbLaps$ executions $\times$ $11$ \exploitationCoefficient $\times$ $31$ seeds = 1 091 200 executions.
The raw data of this evaluation is publicly available on GitHub.\footnote{The raw evaluation data of \banditrepair: \url{https://goo.gl/TJezRr}.}
We answer to all research questions based on this data.

\subsubsection{Validity of Repair Decision Sequences}
\label{sec:validity}
For a given decision sequence taken in response to a failure, we assess its validity according to the laps oracle. 
In those experiments, the laps oracle are directly extracted from the test case reproducing the field failure.
As such, a decision sequence is considered as valid, if no null pointer exception is thrown, and no other exception is thrown.
A decision sequence is considered as invalid, if the original null pointer exception is thrown (meaning that that there is no possible decision at the failure point), or another exception is thrown and not caught.
When the test case contains domain-specific assertions beyond the occurrence or not of exceptions, we keep them, and a decision sequence is considered valid if all assertions pass after the application of the runtime patch. This is the case for 14/\nbBugs failures.

\begin{figure}
\centering
\caption{Excerpt of the decision tree of Math-988A. One path is this tree is a ``decision sequence'', one path resulting in a success (OK) is a ``runtime patch''. DPx refers to a decision point upon a null dereference failure.}
\label{fig:decisionTree} 
\includegraphics[width=0.45\textwidth]{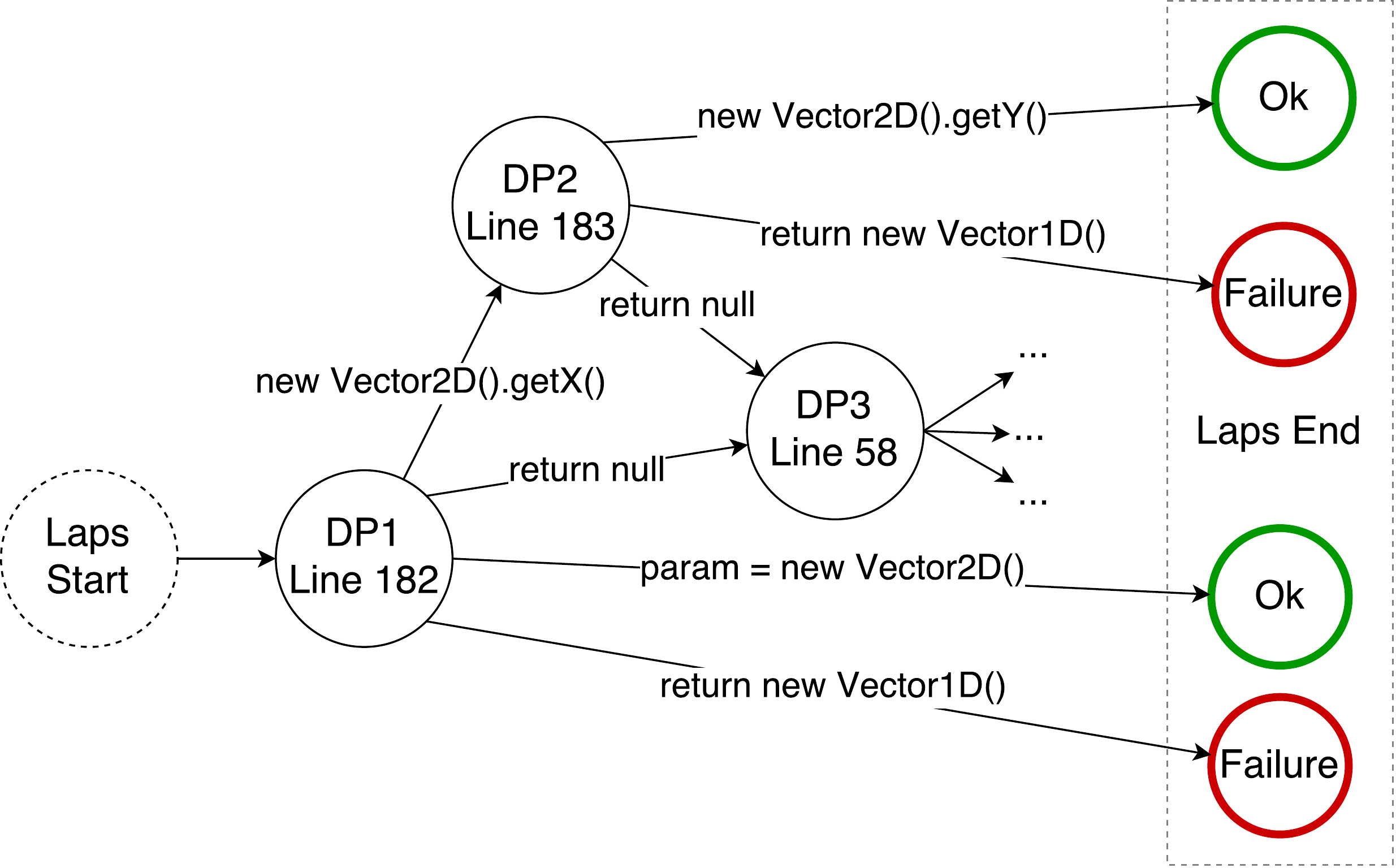}
\end{figure}

\subsection{Case Study}
\label{sec:case-study}

By applying the protocol of \autoref{sec:protocol}, we obtain runs for which runtime patches are identified.
We now discuss the runtime patches of  Math-988A, where the null pointer exception is thrown during the geometrical computation of the intersection of two lines when they do not have intersection.
The initial null pointer exception is triggered in the return statement of method ``toSubSpace'' which returns an object of type Vector1D.
The null pointer exception appears when a Vector2D parameter is null and methods getX and getY are called on it.
As shown in \autoref{tab:repair_space} the size of the decision sequences varies between 1 and 3 for this bug, meaning that there are between 1 and 3 null dereferences happening depending on the selected decisions, for which a decision is taken.

Overall, \banditrepair identifies the following runtime patches:
\begin{enumerate*}
\item initialize the null parameter with a new instance (1 decision point)
\item use a new and disposable instance of Vector2D at the both places where the null parameter is used (between 2 decision points)
\item return null either at the first NPE location or at the second one, triggering another decision in the caller (between 2 and 3 decision points)
\item return a new instance Vector1D (1 decision point)
\end{enumerate*}
Recall that in our setup, the laps begins with the execution of the test that reproduces the field failure and it stops at the end of the test execution. 
In this case, the test contains Junit assertions checking the expected correct behavior (which is to return null when no intersection exists). The runtime patch passes those assertions, it means that in this case, runtime repair achieves full correctness. By comparing the decision sequence to the human patch, they are indeed equivalent, yet different. This confirms that there sometimes exists multiple execution paths for achieving the same computational effect.
\autoref{fig:decisionTree} shows an excerpt of the possible paths in the decision tree from laps start to laps oracle evaluation. 

\subsection{Results}

\subsubsection{\rqCore}
\label{sec:rq1}

\begin{table}[t]
\caption{Data set of \nbBugs bugs with null dereference in six Apache open-source projects.}
\label{tab:repair_space}
\centering
\resizebox{0.40\textwidth}{!}{
\setlength\tabcolsep{0.7 ex}
\begin{tabular}{|l|r|r|r|r|r|r|r|r|r|}\hline
\multirow{6}{*}{Bug ID}     & \multirow{6}{*}{\rotatebox{90}{\tabincell{c}{Nb encountered \\ decision points}}}
                            & \multirow{6}{*}{\rotatebox{90}{\tabincell{c}{Nb \\ decision seq.}}} 
                            & \multirow{6}{*}{\rotatebox{90}{\tabincell{c}{Nb valid \\ runtime patches}}}
                            & \multicolumn{3}{c|}{\multirow{5}{*}{\tabincell{c}{$\mid$ Valid \\ runtime patches.$\mid$}}} \\
                            & & & & \multicolumn{3}{l|}{} \\
                            & & & & \multicolumn{3}{l|}{} \\
                            & & & & \multicolumn{3}{l|}{} \\
                            & & & & \multicolumn{3}{l|}{} \\\cline{5-7}
                            & & & &  Min. & Med. & Max.  \\
                            \hline
Collections-360 &      2 &     45 &     16 &      2 &      2 &      2  \\
Felix-4960      &      1 &     10 &      4 &      1 &      1 &      1  \\
Lang-304        &      1 &      7 &      6 &      1 &      1 &      1  \\
Lang-587        &      1 &     28 &      1 &      1 &      1 &      1  \\
Lang-703        &      4 &    459 &    130 &      2 &      2 &      2  \\
Math-1115       &      1 &      5 &      5 &      1 &      1 &      1  \\
Math-1117       &     21 &  51785 &   7708 &      7 &      8 &      8  \\
Math-290        &      1 &     14 &      4 &      1 &      1 &      1  \\
Math-305        &      1 &      4 &      3 &      1 &      1 &      1  \\
Math-369        &      2 &     14 &      0 &  ---   &  ---   &  ---    \\
Math-988A       &      3 &    576 &    383 &      1 &      2 &      3  \\
Math-988B       &      1 &     32 &     17 &      1 &      1 &      1  \\
Pdfbox-2812     &      8 &    294 &    168 &      1 &      6 &      7  \\
Pdfbox-2965     &      1 &      4 &      3 &      1 &      1 &      1  \\
Pdfbox-2995     &      1 &      5 &      1 &      1 &      1 &      1  \\
Sling-4982      &      2 &     16 &     11 &      1 &      1 &      1  \\
\hline
Total           &     51 &  53298 &   8460 &      1 &      1 &      8  \\
\hline
\end{tabular}
}
\end{table}

We analyze the data obtained with the experiment described in \autoref{sec:exhaustive-exploration}, consisting of exhaustively exploring the search space of runtime patches for null dereferences.
We create a table that contains the core metrics we are interested in, it is reproduced in \autoref{tab:repair_space}.  

\autoref{tab:repair_space} reads as follows.
Each line corresponds to a failure of our dataset. Each column gives the value of a metric of interest. 
The first column contains the name of each bug.
The third column contains the number of possible repair decision sequences for this failure.
The fourth column contains the number of runtime patches (valid decision sequences for which the laps oracle has stated that the decision sequence has worked).
The fifth column contains the minimum/median/maximum number of decisions taken for valid decision sequences.

For example the first line of \autoref{tab:repair_space} details the bug Collections-360.
To repair this failure at runtime, there are two possible decision points, which, when they are systematically unfolded, correspond to 45 possible decision sequences, 16 of which being valid according to the laps oracle.
The size of the valid decision sequences is always equals to $2$, which means that there must be two decisions taken in a row to handle the failure. 

The core assumption of \banditrepair is that there exists multiple alternative decisions to repair a failure at runtime. 
This assumption is reflected by the number of explored decision sequences, which is exactly the size of our search space since we conduct an exhaustive exploration. 
In this experiment, it ranges between 4 decisions (for Math-305 and PdfBox-2965) to 576 for Math-988A and 53951 for Math-1117.

Overall, we notice a great variance of the size of the repair space.
To sum up, for all \nbBugs failures of our benchmark, there exists alternative repair decisions to be taken at runtime.
However, this was not at all an inclusion criterion for building the benchmark.
It strongly suggests that alternative runtime repair decisions are prevalent for null dereference failures, and validates our core assumption.

We also see in \autoref{tab:repair_space} that there is a correlation between the number of activated decision points for a given failure and the number of possible decision sequences. 
For instance, for Felix-4960, there is only one activated decision point (at the failure point where the null pointer exception is about to happen), and 10 possible decisions can be taken at this point. 
On the contrary, Math-1117 has the biggest number of activated decision points resulting in a huge search space, 51 785 decision sequences.
This correlation is expected and explained analytically as follows.
Once a first decision is made at the failure point (where the null dereference is about to happen), many alternative execution paths are uncovered. Then, a combinatorial explosion of stacked decisions happens. If we assume that there are 5 alternative decisions at the first decision point, and that each of them triggers a different execution path and another decision point (all different) with 10 alternatives, it directly results in $5\times 10=50$ possible decision sequences. One can easily extrapolate that for more than 2 stacked decision points, there is a combinatorial explosion.
In general, the size of the repair space depends on:
\begin{enumerate*}
\item the overall number of decision points activated for a given failure
\item the number of possible decisions at each decision point
\item and the correlation between each decision point, that is the extent to which one decision influences the number of possible subsequent decisions to be taken.
\end{enumerate*}
For failures with large \# of explored decision sequences, it means that runtime repair unfolds a large number of diverse program states and their corresponding subsequent executions.

\answer{1}{The core assumption of \banditrepair holds: there are multiple alternative decision sequences to handle null dereferences. 
\banditrepair draws a precise picture of the runtime repair search space. In our experiment, there are 11/\nbBugs failures for which we observe more than 10 possible decision sequences (column ``Nb decision seq.'') for the same failure and according to our runtime patch model, with a maximum value of 51785 (for Math-1117)}

\subsubsection{\rqProportionValid}
\label{sec:rq2}

Now that we have a clearer picture of the size of the search space, we are interested in knowing whether there exists multiple valid decision sequences in that space.
To do so, we still consider the exhaustive study protocol described in  \autoref{sec:exhaustive-exploration} whose results are give in  \autoref{tab:repair_space}.
We concentrate especially on the column showing the number of valid decision sequences. We compare it against the column representing the size of the search space, i.e. the total number of possible decision sequences. 
For instance, for Collections-360, the search space contains 45 possible decision sequences, in which 16 are valid according to the laps oracle (the absence of null pointer exception and the two  assertions at the end of the test case reproducing the failure pass).
This makes a proportion of $16/45=36\%$ of valid decision sequences in the search space.

We notice several interesting extremum cases in \autoref{tab:repair_space}.
First there are two failures -- Lang-587 and PdfBox-2995 -- for which only 1 valid decision sequence exists.
Also there is one failure for which all decisions remove the failures, this is Math-1115 for which 5 out of 5 possible decision sequences are valid.
In general there is a great diversity of fertility (the proportion of runtime patches in the search space), which can be explained by two factors: first the strength of the laps oracle, which is in our case the strength of the assertions at the end of the test case that reproduces the failure (beyond not throwing a null pointer exception).
The second factor is related to the new program states that are explored, once a first decision has been taken.
If those speculative program states are too unrealistic, there is a great chance that the corresponding decision sequences are invalid (for instance because another exception is thrown).
Along the same line, if the first decision taken, at the failure point, yields exotic program states, it is unlikely that the subsequent decisions put the system back into a viable state.

\answer{2}{In our benchmark, the proportion of valid decision sequences varies from 0/14\% to 100\%, from 0 to 7708/51785 valid runtime patches. This great variation is due to the varying complexity of the system state at the failure point, and the hardness of the laps oracle. 
When the proportion of valid repair decision sequences is high, it means that \banditrepair is able to quickly find a valid runtime patch, based on a small number of failure occurrences.}

\begin{table}
\caption{Core Metrics of Running \banditrepair on the \nbBugs Failures of our Benchmark.
Column ``Nb valid runtime patches'' shows that more than 1 sequence of repair decisions which is valid exists.}
\label{tab:results}
\centering
\resizebox*{!}{0.9\textheight}{
\setlength\tabcolsep{2.5 ex}
\begin{tabular}{|l|c|r|r|r|r|r|r|r|r|r|r|r|r|r|r|}\hline
Bug ID                      & \exploitationCoefficient
                            & \rotatebox{90}{\tabincell{c}{Nb activated \\ decision points}}
                            & \rotatebox{90}{\tabincell{c}{Nb explored \\ decision seq.}}
                            & \rotatebox{90}{\tabincell{c}{Nb valid \\ runtime patches}}
                            & \rotatebox{90}{\tabincell{c}{\% fixed \\ failures}}
                            & \rotatebox{90}{\tabincell{c}{Nb laps before \\ max exploration}} \\
                            \hline
\multirow{4}{*}{Collections-360}
 & $1.0$   &      2 &      3 &      1 & 99\% &      4  \\
 & $0.8$   &      2 &     27 &     13 & 77\% &     45  \\
 & $0.2$   &      2 &     38 &     15 & 31\% &     22  \\
 & $0.0$   &      2 &     38 &     15 & 20\% &     20  \\
\hline
\multirow{4}{*}{Felix-4960}
 & $1.0$   &      1 &      2 &      1 & 100\% &      3  \\
 & $0.8$   &      1 &     10 &      4 & 88\% &     36  \\
 & $0.2$   &      1 &     10 &      4 & 52\% &     13  \\
 & $0.0$   &      1 &     10 &      4 & 40\% &     11  \\
\hline
\multirow{4}{*}{Lang-304}
 & $1.0$   &      1 &      1 &      1 & 100\% &      2  \\
 & $0.8$   &      1 &      7 &      6 & 96\% &     31  \\
 & $0.2$   &      1 &      7 &      6 & 87\% &     12  \\
 & $0.0$   &      1 &      7 &      6 & 84\% &     11  \\
\hline
\multirow{4}{*}{Lang-587}
 & $1.0$   &      1 &     13 &      1 & 92\% &     19  \\
 & $0.8$   &      1 &     28 &      1 & 76\% &     96  \\
 & $0.2$   &      1 &     28 &      1 & 22\% &     33  \\
 & $0.0$   &      1 &     28 &      1 & 4\% &     29  \\
\hline
\multirow{4}{*}{Lang-703}
 & $1.0$   &      4 &     10 &      2 & 96\% &     14  \\
 & $0.8$   &      4 &     83 &     37 & 76\% &    199  \\
 & $0.2$   &      4 &    139 &     43 & 28\% &     97  \\
 & $0.0$   &      4 &    130 &     32 & 18\% &     75  \\
\hline
\multirow{4}{*}{Math-1115}
 & $1.0$   &      1 &      1 &      1 & 100\% &      2  \\
 & $0.8$   &      1 &      5 &      5 & 100\% &     22  \\
 & $0.2$   &      1 &      5 &      5 & 100\% &      7  \\
 & $0.0$   &      1 &      5 &      5 & 100\% &      6  \\
\hline
\multirow{4}{*}{Math-1117}
 & $1.0$   &     10 &     79 &      0 & 0\% &    192  \\
 & $0.8$   &     10 &     79 &      0 & 0\% &    190  \\
 & $0.2$   &     10 &     81 &      0 & 0\% &    186  \\
 & $0.0$   &     10 &     78 &      0 & 0\% &    177  \\
\hline
\multirow{4}{*}{Math-290}
 & $1.0$   &      1 &      3 &      1 & 99\% &      4  \\
 & $0.8$   &      1 &     14 &      4 & 84\% &     54  \\
 & $0.2$   &      1 &     14 &      4 & 42\% &     18  \\
 & $0.0$   &      1 &     14 &      4 & 28\% &     15  \\
\hline
\multirow{4}{*}{Math-305}
 & $1.0$   &      1 &      1 &      1 & 100\% &      2  \\
 & $0.8$   &      1 &      4 &      3 & 94\% &     15  \\
 & $0.2$   &      1 &      4 &      3 & 80\% &      6  \\
 & $0.0$   &      1 &      4 &      3 & 75\% &      5  \\
\hline
\multirow{4}{*}{Math-369}
 & $1.0$   &      2 &     13 &      0 & 0\% &    101  \\
 & $0.8$   &      2 &     13 &      0 & 0\% &     81  \\
 & $0.2$   &      2 &     13 &      0 & 0\% &     49  \\
 & $0.0$   &      2 &     13 &      0 & 0\% &     49  \\
\hline
\multirow{4}{*}{Math-988A}
 & $1.0$   &      2 &      1 &      1 & 100\% &      2  \\
 & $0.8$   &      3 &     92 &     86 & 98\% &    129  \\
 & $0.2$   &      3 &    114 &    104 & 94\% &    181  \\
 & $0.0$   &      3 &    117 &    105 & 92\% &    190  \\
\hline
\multirow{4}{*}{Math-988B}
 & $1.0$   &      1 &      2 &      1 & 100\% &      3  \\
 & $0.8$   &      1 &     32 &     17 & 90\% &    146  \\
 & $0.2$   &      1 &     32 &     17 & 62\% &     40  \\
 & $0.0$   &      1 &     32 &     17 & 52\% &     33  \\
\hline
\multirow{4}{*}{Pdfbox-2812}
 & $1.0$   &      2 &      1 &      1 & 100\% &      2  \\
 & $0.8$   &      8 &     23 &     19 & 94\% &    127  \\
 & $0.2$   &      8 &     43 &     33 & 74\% &     64  \\
 & $0.0$   &      8 &     49 &     36 & 64\% &     55  \\
\hline
\multirow{4}{*}{Pdfbox-2965}
 & $1.0$   &      1 &      1 &      1 & 100\% &      2  \\
 & $0.8$   &      1 &      4 &      3 & 95\% &     19  \\
 & $0.2$   &      1 &      4 &      3 & 80\% &      6  \\
 & $0.0$   &      1 &      4 &      3 & 74\% &      5  \\
\hline
\multirow{4}{*}{Pdfbox-2995}
 & $1.0$   &      1 &      3 &      1 & 99\% &      4  \\
 & $0.8$   &      1 &      5 &      1 & 82\% &     15  \\
 & $0.2$   &      1 &      5 &      1 & 36\% &      6  \\
 & $0.0$   &      1 &      5 &      1 & 20\% &      6  \\
\hline
\multirow{4}{*}{Sling-4982}
 & $1.0$   &      2 &      3 &      2 & 100\% &      4  \\
 & $0.8$   &      2 &     16 &     11 & 93\% &     76  \\
 & $0.2$   &      2 &     16 &     11 & 75\% &     22  \\
 & $0.0$   &      2 &     16 &     11 & 68\% &     17  \\
\hline
\hline
\multirow{3}{*}{Average}
 & $1.0$   &   2.06 &   8.56 &      1 & 86\% &  22.50 \\
 & $0.8$   &   2.50 &  27.62 &  13.12 & 77\% &  80.06 \\
 & $0.2$   &   2.50 &  34.56 &  15.62 & 53\% &  47.62 \\
 & $0.0$   &   2.50 &  34.38 &  15.19 & 46\% &     44 \\
\hline

\end{tabular}
}
\end{table}

\subsubsection{\rqBetterDecision}
\label{sec:rq3}

We have shown in {\bf RQ2} that there are multiple valid runtime patches.
Now, we aim to determine which runtime patches from our runtime patch portfolio are better  with respect to their size, as measured by the number of decisions.
To do so, we study the results of the exhaustive study protocol described in  \autoref{sec:exhaustive-exploration} whose results are given in \autoref{tab:repair_space}.
We especially concentrate on the column showing the size of the valid decision sequences.
This column gives the minimum, median and maximum size within the portfolio. 
For instance, for PDFBox-2812, the minimal size in number of decisions among all runtime patches is $1$, the median size is $6$ and the maximal size is $7$.

This data supports the following findings.
First, one sees that there exists runtime patches composed of more than one decision. For instance, for Collections-360, all runtime patches contains 2 decisions. 
Since our failure and runtime repair model is specific to null pointer exceptions, it means that there exists decisions for which the null dereference problem is not definitely solved by the first decision, and that another null dereference happens later. This is indeed the case for Collections-360 where the null variable is used twice in two different methods.

Second, in $10/\nbBugs$ of our dataset, the runtime patches are always composed of a single decision. This is strongly correlated to the size of the search space (third column, \# of decision sequences), hence indicating that the test case reproducing the production failure, sets up a program state that is repairable in one shot.

Third, in $3/\nbBugs$ failures, there are runtime patches of different size (Math-988A, PDFBox-2812, Math-1117). For instance, for Math-988A, there exists runtime patches of 1, 2 and 3 decisions. This means that there are decisions at the failure point that definitely solve the problem according to the laps oracle (those of patches of size 1). Assuming that the search first finds a complex runtime patch with many decisions, it is indeed necessary to further explore the search space in order to identify a smaller, hence better runtime patch.
Forth, in one case (Math-369), there are several decision taken, but none of them are valid and there is no runtime patch. All decision sequences are invalidated by the laps oracle (the assertions of the test case reproducing the field failure).

When a runtime patch of size 1 is found, it may be argued that it is best, and that the speculative exploration could be stopped. This is not what \banditrepair does, because it aims at building a portfolio of runtime patches, and there may exists other runtime patches of size 1 in the search space.

\answer{3}{For $5/\nbBugs$ failures of our benchmark, the search space contains composite runtime patches that have more than one decision. For $3/\nbBugs$ failures, the possible runtime patches have disparate sizes, and exploratory search of new runtime patches enables to find smaller runtime patches.}

\subsubsection{\rqExplorationCoefficient}

Following the protocol based on bandit exploration described in \autoref{sec:bandit-exploration}, we vary the value of the exploitation coefficient \exploitationCoefficient and explore the impact it has on the search process.
The results are given in \autoref{tab:results}.

\autoref{tab:results} reads as follows.
Each line corresponds to a reproduced failure of our dataset. Each column gives the value of a metric of interest. 
Each failure line is split in four, corresponding to four different exploitation coefficient \exploitationCoefficient (0, 0.2, 0.8 and 1).
The first column contains the name of each bug.
The second column contains the value of the \exploitationCoefficient parameter for a given line (as defined in \autoref{sec:decision-taking}).
The third column is the number of encountered decision points over the \nbLaps laps.
The fourth column contains the number of explored decision sequences.
The fifth column contains the number of valid decision sequences.
The sixth column contains the number of laps before a decision sequence is valid and succeeds the laps oracle.

For example, let us consider Collections-360 with \exploitationCoefficient set to $0.8$ (second line of the four lines for this failure). During these runs, two locations in code trigger a null dereference, which means that two decision points are activated. The combination of decisions over those two decision points results in 27 explored decision sequences for \exploitationCoefficient=1 (this means that the first 26 decision sequences are invalid, and the $27^{th}$ is a runtime patch which is then exploited).
Among those 26 decision sequences, 12 are considered valid according to the laps oracle.
Among the 11 runs, it took a median of 3 runs before finding a valid decision sequence that fixes the failure.

The value of the exploitation coefficient \exploitationCoefficient has an impact on the repair as follows.
First, it has an impact on the number of activated decision points. For Math-1117 and Math-369, Math-998A and Pdfbox-2812, if we explore more and exploit less (lower \exploitationCoefficient), we explore less of the search space, and hence activate fewer decision points. 
The number of activated decision points is a coarse grain view of the explored search space. The fourth column showing the number of explored decision sequences better reflects what we are interested in. For all failures, if we increase the amount of exploration (lower \exploitationCoefficient), this indeed results in trying out more decision sequences (corresponding to a larger figure in 4th column).
This validates the overall tradeoff architecture of \banditrepair for balancing exploration and exploitation.

\begin{figure}
\centering
\includegraphics[width=0.47\textwidth]{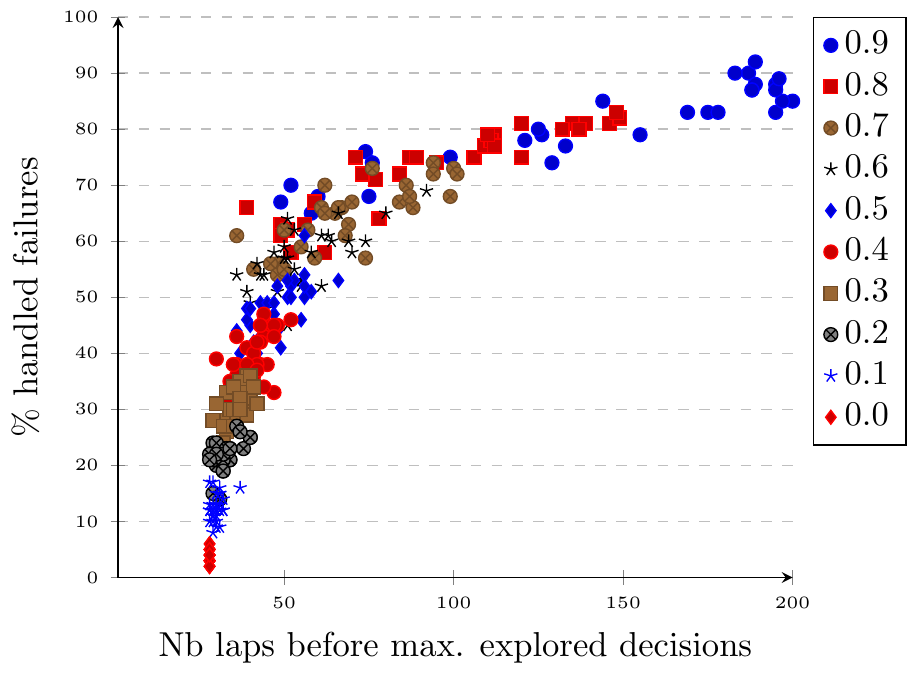}
\caption{Pareto Front of Runtime Repair for Lang-587 (bettered view on screen or with color printing). Bigger exploitation coefficient yields higher proportion of handled failures, but slower time to explore the repair search space.} 
\label{plot:exploitation-lang587}
\end{figure}

Now, let us consider the number of valid decision sequences.
As expected, the proportion is valid sequences is roughly the same as the ones found during exhaustive exploration.
Also, when there is full exploitation (\exploitationCoefficient$=1$), the search stops when a valid decision sequence is found, which can be observed in the table (only one runtime patch is identified for the columns where \exploitationCoefficient$=1$), this is an evidence of the correctness of our implementation. 
Interestingly, for Math-1117, there is no valid decision sequence found (while such sequence exists in the search space as shown in \autoref{tab:results}). The reason is that the number of runs of the experiment (\nbLaps) is too small compared to the size of the search space, and the valid decision sequence has not yet been found after \nbLaps runs.

Now we concentrate on the number of handled failures which directly reflects the number of users affected by the bug.
For instance, for Pdfbox-2812, if one exploits a lot (\exploitationCoefficient=0.8) one handles 189 failures and builds a portfolio of 19 alternative runtime patches.
On the contrary, in one exploits less and explores a lot (\exploitationCoefficient=0.2), one handles 147 failures but collects 33 runtime patches ($>19$).
The rule is the more one exploits (bigger \exploitationCoefficient, such as 1 or 0.8) the more failures are handled. 
However, this is done to the price of not building a portfolio of alternative runtime patches.
This is the essence of the balance between exploration and exploitation.

We now graphically depict the tradeoff between exploration and exploitation of runtime patches. 
\autoref{plot:exploitation-lang587} is scatter plot of evaluation runs for failure Lang-587.
Each dot is a run for a given random seed and exploitation coefficient. Hence, there are $10*31=310$ dots.

The dots are colored by the value of the exploitation coefficient \exploitationCoefficient. 
For a given run (dot), the X axis is the number of laps before the exploration becomes unsuccessful (no new decision sequences discovered after that), it corresponds to the 7th column of \autoref{tab:results}.
The Y-axis corresponds to the proportion of fixed failures shown in the 6th column of \autoref{tab:results}.
For instance, the top-most blue point, is an exploration with \exploitationCoefficient=0.9, for which 180/\nbLaps (90\%) runs are successfully repaired at runtime.
In this figure, one clearly sees the Pareto front of the tradeoff between exploration and exploitation.
The more one exploits, the longer it takes to explore the repair search space, but the more failures are handled.
On the contrary, if one explores a lot (low \exploitationCoefficient), the search space is traversed really fast, building a large portfolio of runtime patches, but with a low proportion of handled failures.
Interestingly, when using \banditrepair, there is irreducible warm up time before finding a valid repair decision sequence: this is the empty space between $x=0$ and $x=28$ at the right hand side of the figure. This is explained by the fact that we explore new decisions in a deterministic manner (only the choice between exploration and exploitation is random) , exploring decisions one after the other in the same order during the exploration phase. 
According to this deterministic order, for Lang-587 shown in this figure, it means that the 28th explored decision sequence is the first valid one.

\answer{4}{The exploitation coefficient \exploitationCoefficient has an impact on the size of the explored search space, the number of repaired failures, and the size of the portfolio of discovered runtime patches. The relation between \exploitationCoefficient and those three core metrics draws a Pareto front of runtime repair.}

\section{Related work}\label{sec:rw}

There are several automatic repair techniques that handle failures at runtime. 

One of the earliest techniques is Ammann and Knight's ``data diversity'' \cite{ammann1988data}, that aims at enabling the computation of a program in the presence of failures.
The idea of data diversity is that, when a failure occurs, the input data is changed so that the new input resulting from the change does not result in the failure.
The assumption is that the output based on this artificial input, through an inverse transformation, remains acceptable in the domain under consideration.
The input transformations can be seen as a kind of runtime patch model. As such, the \banditrepair algorithm could be used to reason on the associated runtime search space.

Demsky et al. \cite{demsky2003automatic} presents a language for the specification of data structure invariants. The invariant specification is used to verify and repair the consistency of data structure instances at runtime. 
The key difference between their work and ours is that \banditrepair is more generic in scope, only requiring a laps model and a laps oracle, which go beyond data structure errors and invariant restoration only.

Rinard et al. \cite{rinard2004enhancing} presents a technique to avoid illegal memory accesses by adding additional code around each memory operation during the compilation process.
For example, the additional code verifies at runtime that the program only uses the allocated memory.
If the memory access is outside the allocated memory, the access is ignored instead crashing with a segmentation fault.
The two differences between this work and \banditrepair are: first \banditrepair can apply different decisions to handle a given failure (and not a single code, hard-coded in the injected code), and second, \banditrepair uses an oracle to reason about the viability of the decision.

Perkins et al. \cite{perkins2009automatically} proposes ClearView, a system for automatically repairing errors in production.
The system consists of monitoring the system execution on low-level registers to learn invariants. 
Those invariants are then monitored, and if a violation of an invariant is detected ClearView forces the restoration.
From an engineering perspective, the difference is \banditrepair reasons on decision sequences, while ClearView analizes each decision in isolation.
From a scientific perspective, our work finely characterizes of the search space and the outcomes of runtime repair based on execution modification. 

Kling et al. \cite{kling2012bolt} propose Bold a system to detect and escape infinite and long-running loops.
On user demand, Bolt is attached to a running application and tries different strategies to escape the infinite loop. 
If a strategy fails, Bolt uses rollback to restore the state of the application and then tries the next strategy.
As \banditrepair, Bolt considers multiple decisions for a given failure, but the main difference is that it does not perform and reason about decision sequences made to handle cascaded errors.

Long et al. \cite{long2014automatic} introduces the idea of “recovery shepherding” in a system called RCV.
Upon certain errors (null dereferences and divide by zero), recovery shepherding consists in returning a manufactured value, as for failure oblivious computing.
The key idea of recovery shepherding is to track the manufactured values so as to see 1) whether they are passed to system calls or files and 2) whether they disappear.
In \banditrepair's runtime patch model, 2/ our 5 kinds of decisions also use manufactured values. However, the key difference is that RCV reasons on each manufactured value in isolation. On the contrary, if an injected manufactured value triggers the creation of another one, what is called ``cascaded errors'' in RCV, \banditrepair will reason on the effect of their combinations (by storing and keeping information about the the actual valid sequence of decision). 

Jula et al. \cite{jula2008deadlock} presents a system to defend against deadlocks at runtime.
The system first detects synchronization patterns of deadlocks, and when the pattern is detected, the system avoids re-occurrences of the deadlock with additional locks.
The pattern detection is related to the detector of instances of the fault model under consideration. However, Jula et al. do not explore and compare alternative locking strategies. We note that the code algorithm of \banditrepair may be plugged on top of their systems to explore the search space of locking sequences.

Hosek and Cadar \cite{hosek2013safe} switch between application versions when a bug is detected.
This technique can handle failures because some bugs disappear while others appear between versions. 
We can also imagine to plug \banditrepair on top of their system to systematically explore the sequences of runtime jumps across versions.

Assure \cite{sidiroglou2009assure} is a self-healing system based on checkpointing and error virtualization.
Error virtualization consists of handling an unknown and unrecoverable error with error handling code that is already present in the system yet designed for handling other errors. 
While Assure does runtime repair by opportunistic reuse of already present recovery code, \banditrepair handles failures by modifying the state or flow according to a runtime patch model.

Carzaniga et al. \cite{carzaniga2010automatic} repair web applications at runtime with set of manually written, API-specific alternatives rules. This set can be seen as a hardcoded set of runtime patches.
On the contrary, \banditrepair does not require a list of alternatives but instead relies on an abstract  runtime patch model that is automatically instantiated at runtime.

Berger and Zorn \cite{berger2006diehard} show that is possible to effectively tolerate
memory errors and provide probabilistic memory safety by randomizing the memory allocation and providing memory replication.
The work by Qin et al. \cite{qin2005safemem} exploits a specific hardware feature called ECC-memory for detecting illegal memory accesses at runtime. The idea of the paper is to use the consistency checks of the ECC-memory to detect illegal memory accesses (for instance due to buffer overflow). 
Both techniques are semantically equivalent in the normal case. On the contrary \banditrepair is meant to reason about the search space of execution modifications that are not semantically equivalent, where one taken decision can impact the rest of the computation. 

Dobolyi and Weimer~\cite{dobolyi2008changing} present a technique to tolerate null dereferences.
Using code transformation, they introduce hooks to a recovery framework.
This framework is responsible for forward recovery of the form of creating a default object of an appropriate type of skipping instructions.
Kent~\cite{kent2008dynamic} proposes alternatives to null pointer exceptions.
He proposes to skip the failure line or exits the method by a return when a null pointer exception is detected.
In those two contributions, there is no reasoning on the search space of runtime repair, as done in \banditrepair. We note that our runtime patch model is inspired by theirs, while richer (method return, variable reuse).

\section{Conclusion}
\label{sec:conclusion}

In this paper, we have presented \banditrepair, a runtime repair system inspired from bandit algorithms in machine learning.
The system explores the search space of runtime repair decisions in a systematic manner. 
As a result, the system controls the trade-off between exploiting known runtime patches that are able to handle a failure and exploring new alternative runtime patches.
We have evaluated the systems with a protocol based on \nbBugs field failures of Java applications, showing that the system uncovers and indeed explores the runtime repair search space. 

This novel and original approach opens new research directions. We are in particular interested in bridging \banditrepair with checkpoint \& rollback in order to perform large-scale parallel speculation execution. Also, we will apply \banditrepair with other runtime patches models: more specific such as arithmetic patch models, and more generic such as catching arbitrary exceptions.  

Our future is to explore varying \exploitationCoefficient over time, as done in sophisticated bandit algorithms, as well as contextual multi-armed bandit where the context is the system state the initial failure point.

\bibliographystyle{abbrvnat}
\bibliography{references}
\balance
\softraggedright

\end{document}